\documentstyle[aps,prl,epsfig]{revtex}

\begin{document}
\draft

\twocolumn[\hsize\textwidth\columnwidth\hsize\csname@twocolumnfalse\endcsname
 
\title{Effective temperature of an aging powder}
\author{Mauro Sellitto
%\inst{1} 
}
\address{
  Laboratoire de Physique, \'Ecole Normale Sup\'erieure de Lyon,\\
  46 All\'ee d'Italie, 69007 Lyon, France.
  }

\maketitle

\begin{abstract}
The aging dynamics and the fluctuation-dissipation relation between the 
spontaneous diffusion induced by a random noise and the drift motion 
induced by a small stirring force are numerically investigated in a 3D 
schematic model of compacting powder: a gravity-driven lattice-gas with 
purely kinetic constraints.
The compaction dynamics is characterized by a super-aging behaviour 
and, in analogy with glasses, 
exhibits a purely dynamical time-scale dependent effective 
temperature. 
A simple experiment to measure this quantity is suggested.
\end{abstract}


\twocolumn\vskip.5pc]\narrowtext

% -----------------------------------------------------------------------
%                            INTRODUCTION 
% -----------------------------------------------------------------------
%
%
{\it Introduction. --} 
Slow relaxation phenomena are ubiquitous in nature.
When one deals with glassy dynamics the challenge is 
the identification of the relevant degrees of freedom  
which makes a thermodynamic description still possible.
In the attempt to provide a unifying framework to the behaviour 
of aging and non-relaxational systems a microscopic definition 
of effective temperature was introduced through a generalized 
fluctuation-dissipation relation~\cite{CuKuPe}. 
This quantity turns out to coincide, at least in mean-field glasses, 
with the Edwards' compactivity~\cite{MoPo,Jorge,FrVi,Theo}, 
previously introduced in a granular matter context~\cite{Sam,Anita}. 
For a class of finite-dimensional and zero-gravity compacting systems, 
recent numerical results have come to support this 
correspondence~\cite{BaKuLoSe}. 
However, in contrast to the case of strong~\cite{Campbell}, or moderately 
strong~\cite{Nowak} vibration regime, 
the absence of a temperature-like quantity in slowly driven compacting
systems under gravity is a rather problematic 
issue~\cite{Nicodemi,BaLo,TaTaVi}. 
In this Rapid Communication we investigate the nature of the aging dynamics 
and the effective temperature in a simple 3D lattice-gas 
model of powder~\cite{SeAr}.
We show that the non-equilibrium dynamics is characterized by a  
three-steps relaxation mechanism and `super-aging' in the
mean-square displacement. 
We then find that during the compaction the response 
to a random perturbation is positive and observe a violation 
of the fluctuation-dissipation relation similar to glasses.
Such features show that it is possible to describe the gravity-driven 
compaction dynamics in terms of a purely dynamical time-scale dependent 
effective temperature.

%\newpage
\bigskip

% -----------------------------------------------------------------------
%                            THE MODEL
% -----------------------------------------------------------------------
%
%
{\it The model. --}
The model we consider was introduced in~\cite{SeAr} as a simple
generalization of the Kob-Andersen model~\cite{KoAn}.
The system consists of $N$ particles on a body centred cubic lattice 
where there can be at most one particle per site. 
There is no cohesion energy among particles and the Hamiltonian is simply
 \begin{eqnarray}
{\cal H}_0 & = & m g \sum_{i=1}^N h_i  \,,
\label{H_0}
\end{eqnarray}
where $g$ is the gravity constant, $h_i$ is the height of the particle $i$, 
and $m$ its mass.
At each time step a particle can move with probability $p$ to a neighboring
empty site if the particle has less than $\nu$ nearest neighbors before 
and after it has moved.
Here $p={\rm min} [1,x^{-\Delta h}]$ where $\Delta h = \pm 1$ is the 
vertical displacement in the attempted elementary move and 
$x = \exp(-mg/T)$ represents the `vibration'.
The kinetic rule is time-reversible and hence the detailed balance is
satisfied.
We therefore assume that statistical properties of the mechanical 
vibrations on the box can be described, after a suitable coarse-graining,
as a thermal bath at temperature $T$.
In the regime of quasi-static flow this assumption, which neglects the 
complicated effects like friction and dissipation between grains,
is a reasonable starting point~\cite{Sam}.
We set throughout the mass $m=1$ and the threshold $\nu=5$. 
Particles are confined in a box closed at both ends and with periodic boundary 
condition in the horizontal direction.
We consider a system of height $16 L$ and transverse surface $L^2$ with 
$L=20$, and number of particles $N=16000$ (corresponding to a global 
density of 0.25).
As shown in~\cite{SeAr}, 
the interplay of kinetic constraints and gravity are enough to reproduce the 
basic aspect of weakly vibrated powder like compaction and segregation 
phenomena, and vibration-dependent asymptotic packing density.
In the following we consider the non-equilibrium features as they show up
in the two-times correlation and response function.

\bigskip

% -----------------------------------------------------------------------
%                            AGING DYNAMICS
% -----------------------------------------------------------------------
%
%
{\it Aging dynamics. --}
Slow relaxation in weakly vibrated powder is closely related to the reduction 
of the free volume available to the particle motion, hence age dependent 
properties are expected~\cite{Struik,BoCuKuMe,Bouchaud}.
The aging dynamics in simple models of granular matter has been recently 
studied by several authors~\cite{NiCo,BaLo,TaTaVi,PhBi}.
For our purposes it can be easily characterized in terms of the `mean-square 
displacement' between two configurations at time 
$t_w$ and $t > t_w$:
\begin{eqnarray}
 B(t,t_w)= \frac{1}{N} \sum_{i=1}^N 
\left\langle \left[h_i(t)-h_i(t_w) \right]^2 
\right\rangle \,,
\label{B}
\end{eqnarray}
where the angular brackets denote the average over the random noise.
The system is initially prepared in a random loose packed state, which 
in this model corresponds to a packing density 
$\rho_{\rm rlp} \simeq 0.707$~\cite{SeAr}, and then the vibration is 
turned on.
The plot of $B(t,t_w)$, see Fig.~\ref{dx02}a, clearly shows the well known
aging effect.
The system does not reach any equilibrium state on the observation 
time-scale, but rather persists in a non-stationary regime:
the particle displacements become slower and slower as the age of the 
system increases.
Interestingly, the Fig.~\ref{dx02}a shows a three steps relaxation mechanism:
a short-time normal diffusion;
an intermediate sub-diffusive regime which is tempting to associate
to the `cage rearrangement';
and finally a relatively faster but still sub-diffusive regime which 
can be figured out as a `cage-diffusion'.
%
%
% -----------------------------------------------------------------------
%                            FIGURE MSD
% -----------------------------------------------------------------------
\begin{figure}
\begin{center}
\epsfig{file=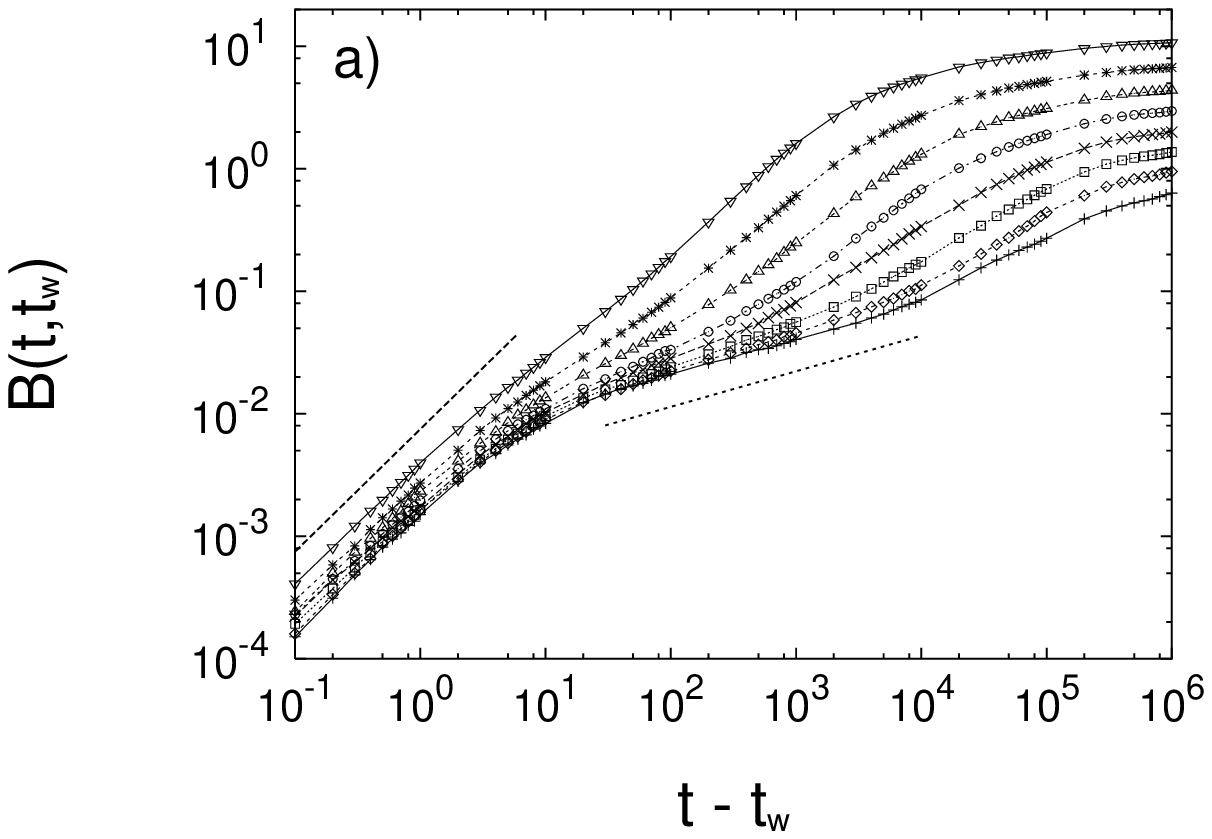,width=7.25cm}
\epsfig{file=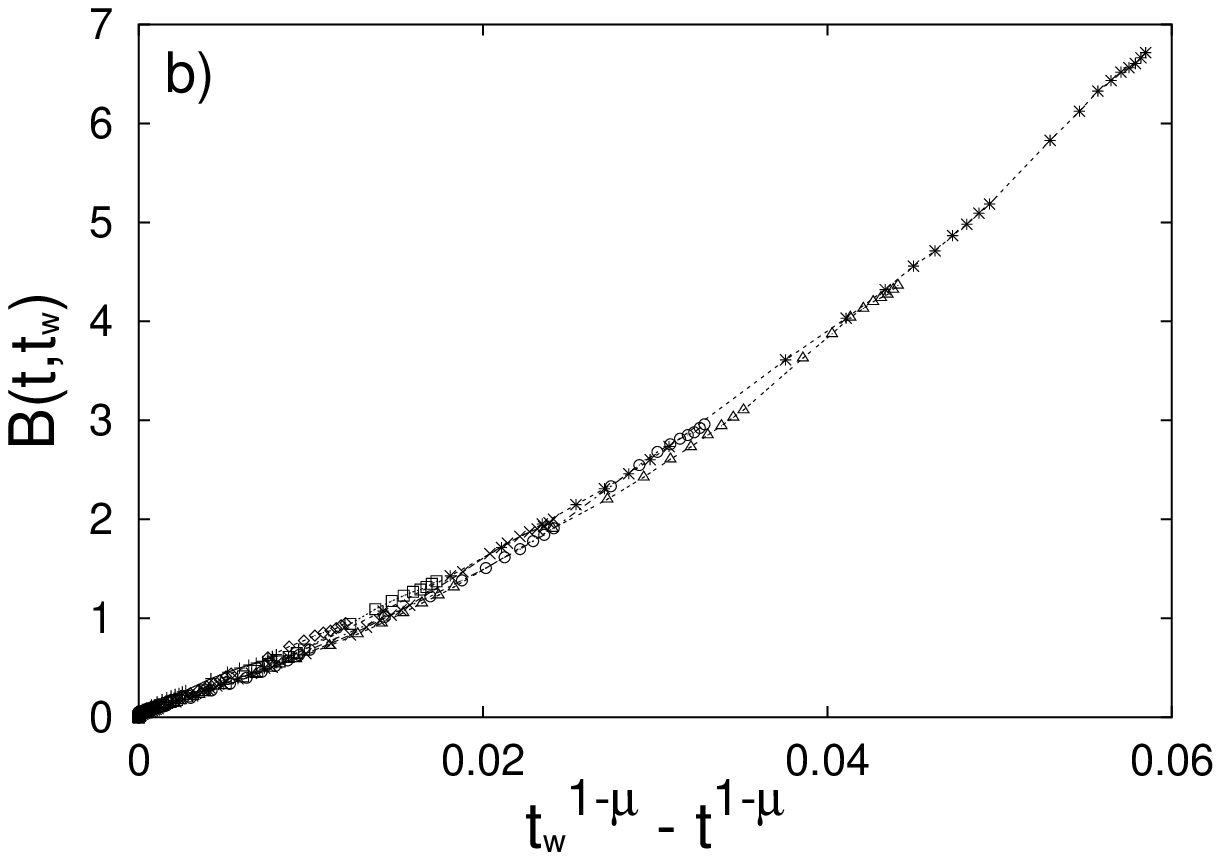,width=7.25cm}
\end{center}
\caption{a) Mean-square displacement $B(t,t_w)$ vs $t-t_w$ 
measured from a random loose packed initial state 
($\rho_{\rm rlp} \simeq 0.707$).
The vibration amplitude is $x=0.2$, and waiting times are 
$t_w=2^k$ with $k=10,...,17$ (from the top to the bottom).
The dashed line represent the short-time normal diffusion regime.
b) The same data plotted vs the scaling variable
$t_w^{1-\mu}-t^{1-\mu}$. 
The super-aging exponent is $\mu \simeq 1.41$ for $x=0.2$.
}
\label{dx02}
\end{figure}    
% -----------------------------------------------------------------------
%
%
Inspired by recent analytic results on a related anomalous diffusion
model~\cite{LeArSe}, we use a scaling function representing 
a super-aging behaviour:
in the Fig.~\ref{dx02}b the mean-square displacement data are plotted vs 
the variable  $t_w^{1-\mu}-t^{1-\mu}$. 
We find that the value $\mu \simeq 1.41$ gives a quite good data 
collapse for vibration $x=0.2$ (Fig.~\ref{dx02}b).
The exponent $\mu$ appears to increase with the vibration, at least
on the accessible time-scale of our simulations.
We obtain for instance  $\mu \simeq 1.28$ for $x=0.05$ and 
$\mu \simeq 1.48$ for $x=0.4$.
According the theory presented in~\cite{LeArSe} the exponent is 
universal and its value is $\mu \simeq 1.48$.
We believe that the discrepancy observed at small $x$ is due to
the fact that the system is still far from the asymptotic 
high packing density regime where the predictions of 
ref.~\cite{LeArSe} apply.
In the super-aging regime, the characteristic relaxation 
time of the system grows faster than its age, a rather 
unusual feature as compared to glassy polymers or random 
magnets~\cite{Bouchaud}.

% -----------------------------------------------------------------------
%                          EFFECTIVE TEMPERATURE
% -----------------------------------------------------------------------
%
%
\bigskip

{\it Effective temperature. --}
The possibility of a thermodynamic description of slow relaxing systems 
is apparently ruled out by the breakdown of time-translation 
invariance, i.e. the presence of aging phenomena.
In granular materials further complications may arise from the presence
of spatial inhomogeneities induced by the boundary conditions and
the gravity direction.
In analogy with glassy systems we show, however, that the violation 
of the fluctuation-dissipation relation is not arbitrary. 
It is such that the degrees of freedom associated with the slow motion can 
be considered as equilibrated  at an effective temperature (vibration) 
higher than the one imposed by the external bath (forcing). 
In order to see this we need to compute the dynamical response function.
We apply to the system a small random stirring force at time $t_w$:
\begin{eqnarray}
{\cal H}_{\epsilon} &=& {\cal H}_0 +
\epsilon \sum_{i=1}^N f_i \, h_i \,\,,
\label{H}
\end{eqnarray}
where $f_i=\pm 1$ independently for each particle.
The linear regime is probed for small enough  values of the perturbation
$\epsilon$.
The integrated response function conjugated to Eq.~(\ref{B}) is
the `staggered displacement'
\begin{eqnarray}
\kappa(t, t_w) &=& \frac{1}{N} \sum_{i=1}^N
\left\langle \, \overline{f_i  \left[ h_i(t) - h_i(t_w) \right]} \,
\right\rangle \,,
\label{kappa}
\end{eqnarray}
where the overline denotes the average over the random stirring force.
At thermal equilibrium $\kappa$ and $B$ are time-translation invariant 
and the Einstein relation holds,
\begin{eqnarray}
\kappa(t-t_w) &=&
\frac{\epsilon}{2T} \, B(t-t_w)  \,.
\label{Einstein}
\end{eqnarray}
It has been suggested that a simple generalization of the previous relation 
in the aging regime provides, in a suitable long-time limit, a reliable 
definition of effective temperature~\cite{CuKuPe}:
\begin{eqnarray}
T_{\rm eff} (t,t_w) &=& \frac{\epsilon}{2} \frac{B(t,t_w)}{\kappa(t,t_w)} \,.
\label{T_eff}
\end{eqnarray}
In a class of solvable mean-field models of glasses~\cite{BoCuKuMe}, 
$T_{\rm eff} (t,t_w)$ has the following properties.
When $t-t_w \sim O(t_w)$,  $T_{\rm eff} (t,t_w)=T$; 
while for $t/t_w \gg O(1)$, $T_{\rm eff} (t,t_w)$ is a constant or 
slowly waiting-time dependent quantity higher than the bath temperature 
$T$.
In recent experiments, effective temperatures
have been measured in glycerol~\cite{GrIs} and laponite~\cite{BeCiLa}.

%
% -----------------------------------------------------------------------
\begin{figure}
\begin{center}
\epsfig{file=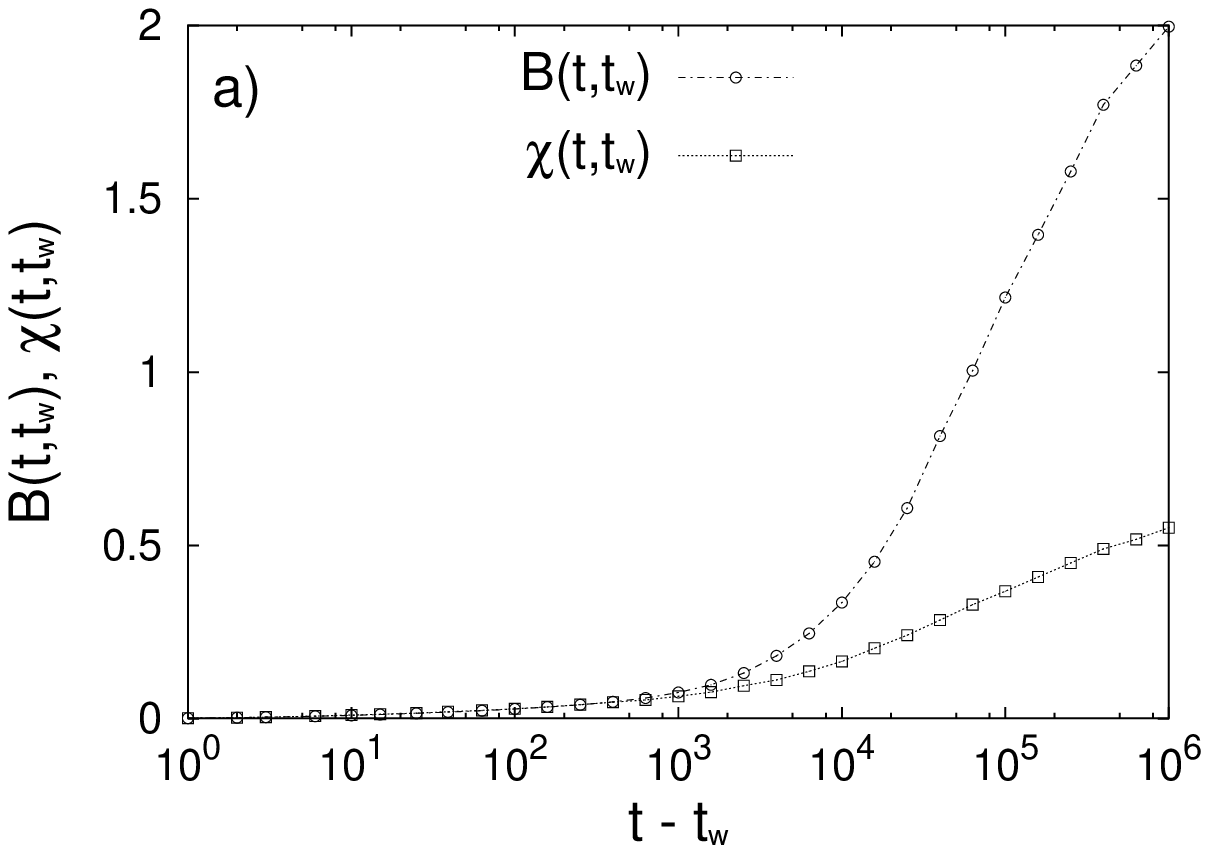,width=7.25cm}
\epsfig{file=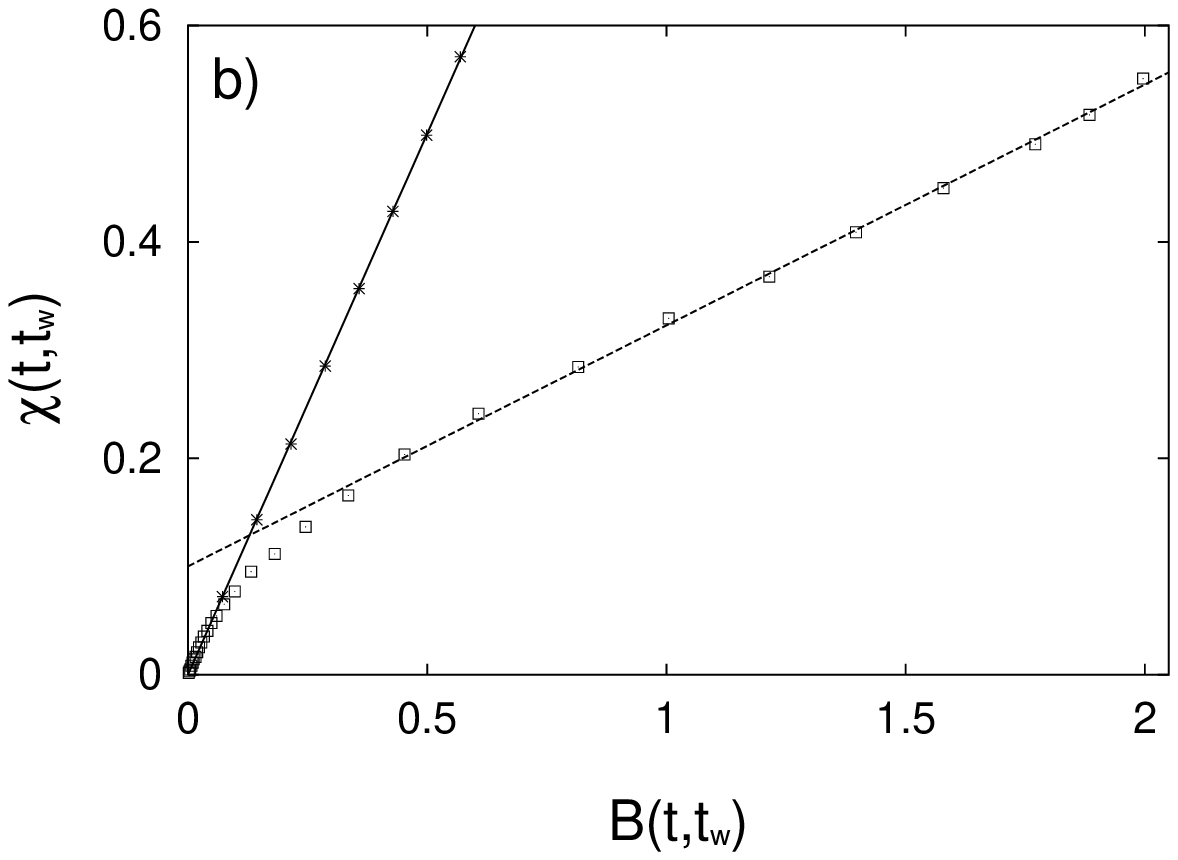,width=7.25cm}
\end{center}
\caption{Measuring the non-equilibrium fluctuation dissipation relation in a 
compaction experiment.
The system is prepared in a random loose packed state, 
$\rho_{\rm rlp} \simeq 0.707$, and then shaken with a vibration $x=0.2$.
The system compacts for a waiting time $t_w = 2^{14}$ after which a small
stirring force is turned on.
a) Comparison of the mean-square displacement $B(t,t_w)$ and the conjugated 
response function $\chi(t,t_w) = 2 T \, \kappa(t,t_w)/\epsilon$ vs $t-t_w$.
b) Parametric plot of $B(t,t_w)$ vs $\chi(t,t_w)$.
The slope of the dashed line is $ \simeq 0.22$. 
The solid line is the equilibrium fluctuation-dissipation theorem 
which is correctly recovered  in the zero-gravity limit, when particles 
diffuse in the whole box and the system reaches a low-density regime. 
}
\label{X}
\end{figure}    
% -----------------------------------------------------------------------

In order to measure the effective temperature 
the system is first prepared in a random-loose packed state.
Then it compacts for a time $t_w$ under a given vibration, $x$. 
At $t_w$ the random perturbation is applied,
and the staggered displacement between configurations at times $t_w$ 
and $t$ is measured.
Following this method we find a positive  monotonic
response function.
At small time separation the integrated response and correlation are equal
while at later time they depart from each other, with the 
response increasing slower than the corresponding 
correlation (see Fig.~\ref{X}a).
The nature of the violation of the fluctuation-dissipation relation
is best represented in the parametric plot of $\kappa(t,t_w)$ vs 
$B(t,t_w)$.
In Fig.~\ref{X}b is shown such a plot for a  waiting time 
$t_w=2^{14}$ MCs and vibration $x=0.2$ 
In close analogy with glassy systems we observe two 
`quasi-equilibrium' regimes.
At short time the usual equilibrium fluctuation-dissipation relation holds 
and  $T_{\rm eff}(t,t_w)=T$.
At later time a strong violation of the fluctuation-dissipation relation
is observed. In this regime the numerical data can be fitted with an 
excellent approximation by a straight line.
This means that the slope of the parametric plot does not depend on the 
observation time and therefore the  effective temperature may only depend 
on $t_w$. 
A closer inspection of Figs.~\ref{dx02} and ~\ref{X} reveals that
the crossover between the two quasi-equilibrium regimes takes place 
over a time scale corresponding to the `cage-rearrangement' motion.
The picture outlined above holds in the whole range
of weak vibrations $0.05 \le x \le 0.4$ and waiting times  $t_w \le 10^5$
that we have explored.
In particular, we find that the effective temperature decreases
with the external vibration (bath temperature) and slowly with 
the waiting time $t_w$.

Finally, to clarify some problems raised in~\cite{Nicodemi},
and discussed in detail in~\cite{BaLo},
we have also studied the response function to a {\it uniform},
either positive or negative, force field.
In both cases we find a non monotonic response function with a negative 
component.
In particular, when kinetic constraints are removed and the system is
at equilibrium, the thermal bath temperature is not recovered with this 
method.
This shows the importance of using stochastic perturbations to define the 
temperature from the fluctuation-dissipation relation.

\bigskip

% -----------------------------------------------------------------------
%                            CONCLUSION
% -----------------------------------------------------------------------
%
%
{\it Conclusion. --}
The notion of effective temperature has been often invoked as the first step 
towards the formulation of a non-equilibrium statistical mechanics for 
systems as different as 
vibrated powders~\cite{Sam,Anita}, 
turbulent fluids~\cite{HoSr}, 
and structural glasses~\cite{CuKuPe,Theo}.
We have shown in this paper that a purely dynamical time-scale dependent 
effective temperature appears during the compaction dynamics of 3D 
constrained lattice-gas models of aging powder.
Mean-field glassy models suggest that such a feature is robust with 
respect to non-relaxational perturbations~\cite{CuKuPe} and time-dependent 
driving forces~\cite{BeCuIg}.
Preliminary results confirm this expectation in more realistic, Hertz 
contact mechanics based models~\cite{Hernan}.

It is worth noting that the perturbed Hamiltonian (\ref{H}) describes a 
particle system with a random distribution of mass with average $m$ and 
standard deviation proportional to $\epsilon$. 
Hence the measure of the effective temperature presented here could be
experimentally carried out in a suitably prepared sample of glass beads, 
provided that $\epsilon$ is small enough to probe the linear response 
regime, and that mass-segregation effects are negligible.
With the particle tracking  experimental facilities (such as, for 
instance, PEPT~\cite{PEPT}) this measure might not be out of reach.            

\bigskip
\bigskip

The author wishes to thank M. Telo da Gama for the kind hospitality at 
{\it Centro de Fisica da Materia Condensada} in Lisbon where this work was 
started.
The support of the EC (contract ERBFMBICT983561) 
is gratefully acknowledged.

% -----------------------------------------------------------------------
%                           BIBLIOGRAPHY  
% -----------------------------------------------------------------------

% -----------------------------------------------------------------------

\begin{thebibliography}{99}

\bibitem{CuKuPe} 
  L.F. Cugliandolo, J. Kurchan and L. Peliti 
  {\it Phys. Rev. E} {\bf 55} 3898 (1997).

\bibitem{MoPo} 
  R. Monasson  and O. Pouliquen 
  {\it Physica A} {\bf 236} 395 (1997).

\bibitem{Jorge} 
  J.~Kurchan, in
  {\it Jamming and Rheology: Constrained Dynamics on Microscopic and Macroscopic Scales}, 
  A.~Liu and S.R.~Nagel~eds. (Taylor \& Francis, New-York, 2001).

\bibitem{FrVi}  
  S. Franz  and M.A. Virasoro
  {\it J. Phys. A} {\bf 33} 891 (2000).

\bibitem{Theo}  
  Th.M. Nieuwenhuizen,
  {\it Phys. Rev. E} {\bf 61} 267 (2000).

\bibitem{Sam}  S.F. Edwards, in   
  {\it Granular Matter}
  Anita Mehta~ed. (Springer-Verlag, Berlin) 1994.

\bibitem{Anita}
  A. Mehta, R.J. Needs, S. Dattagupta,
  {\it J. Stat. Phys.} {\bf 68},  1131 (1992).      

\bibitem{BaKuLoSe} 
  A. Barrat, J. Kurchan, V. Loreto and M. Sellitto,
  {\it Phys. Rev. Lett.} {\bf 85} 5034 (2000); 
  {\it Phys. Rev. E} {\bf 63} 051301 (2001). 

\bibitem{Campbell}
  C.S. Campbell
  {\it Annu. Rev.  Fluid Mech.} {\bf 22} 57 (1990).
  
\bibitem{Nowak} 
  E.R. Nowak, J.B. Knight, E. Ben-Naim, H.M. Jaeger and S. Nagel 
  {\it Phys. Rev. E} {\bf 57} 1971 (1998).

\bibitem{Nicodemi} 
  M. Nicodemi,
  {\it Phys. Rev. Lett.} {\bf 82} 3734 (1999).
 
\bibitem{BaLo} 
  A. Barrat and V. Loreto,
  {\it J. Phys. A} {\bf 33} 4401 (2000).

\bibitem{TaTaVi}
  J. Talbot, G. Tarjus and P. Viot,
  {\tt cond-mat/0008183}.

\bibitem{SeAr}
  M. Sellitto and J.J. Arenzon,  
  {\it Phys. Rev. E} {\bf 62}  7793 (2000).

\bibitem{KoAn} 
  W. Kob and H.C. Andersen 
  {\it Phys. Rev. E} {\bf 48}  4364 (1993).

\bibitem{Struik} 
  L.C.E. Struik,
  {\it Physical Aging in Amorphous Polymers and Other Materials} 
  (Elsevier, Amsterdam, 1978) Chap.~7.

\bibitem{BoCuKuMe}
  J.-P.  Bouchaud, L.F. Cugliandolo, J. Kurchan  and M. M\'ezard,
  in {\it Spin-glasses and random fields}, A.P. Young ed.
  (World Scientific, Singapore) 1998.
 
\bibitem{Bouchaud}
  J.-P.  Bouchaud, 
  in  {\it Soft and Fragile Matter}
  M.E.~Cates, and M.R.~Evans~eds. (I$o$P, Bristol) 2000.

\bibitem{NiCo}
  M. Nicodemi and A. Coniglio
  {\it Phys. Rev. Lett.} {\bf 82} 916 (1999). 

\bibitem{PhBi} P. Philippe and D. Bideau,
  {\it Phys. Rev. E} {\bf 63} 051304 (2001). 

\bibitem{LeArSe}
  Y. Levin, J.J. Arenzon, and M. Sellitto, 
  {\tt cond-mat/0012041}.
 
\bibitem{GrIs} 
  T.S. Grigera and N.E. Israeloff
  {\it Phys. Rev. Lett.} {\bf 83} 5038 (1999). 

\bibitem{BeCiLa} 
 L. Bellon, S. Ciliberto and C. Laroche, 
  {\it Europhys. Lett.} {\bf 53} 511 (2001). 

\bibitem{HoSr} 
  P. Hohenberg and B. Shraiman,
  {\it  Physica D} {\bf 37} 109 (1989).


\bibitem{BeCuIg}
  L. Berthier, L.F. Cugliandolo, and J.L. Iguain,
  {\it Phys. Rev. E} {\bf 63} 051302 (2001). 
           

\bibitem{Hernan} H. Makse, private communication.

\bibitem{PEPT}
  R.D. Wildman, J.H. Huntley, J.-P. Hansen, D.J. Parker, and D.A. Allen
  {\it Phys. Rev. E} {\bf 62} 3826 (2000).              
                                 
 
\end{thebibliography}
\end{document}